\documentclass[journal,10pt]{IEEEtran}
\usepackage{cite}
\usepackage{graphicx}
\usepackage{enumitem}
\setlist[itemize]{leftmargin=*}
\usepackage{epstopdf}
\usepackage[cmex10]{amsmath}\interdisplaylinepenalty=2500
\usepackage[caption=false]{subfig}
\usepackage{relsize}
\usepackage{nicefrac}
\usepackage{amssymb,bm,upgreek}
\usepackage{algorithm}
\usepackage{algpseudocode}
\usepackage{multicol}
\usepackage{xcolor}
\newtheorem{theorem}{Theorem}
\newtheorem{lemma}{Lemma}

\newtheorem{remark}{Remark}

\begin{document}

\title{Location-Aware Pilot Allocation in Multi-Cell Multi-User Massive MIMO Networks}

\author{\IEEEauthorblockN{Noman Akbar,~\IEEEmembership{Student Member,~IEEE}, Shihao Yan,~\IEEEmembership{Member,~IEEE}, Nan Yang,~\IEEEmembership{Member,~IEEE}, and \\ Jinhong Yuan,~\IEEEmembership{Fellow,~IEEE}}
\thanks{Copyright {\textcopyright} 2015 IEEE. Personal use of this material is permitted. However, permission to use this material for any other purposes must be obtained from the IEEE by sending a request to pubs-permissions@ieee.org.}
\thanks{N.~Akbar and N.~Yang are with the Research School of Engineering, Australian National University, Canberra, Australia (e-mails: \{noman.akbar, nan.yang,\}@anu.edu.au). S.~Yan is with the School of Engineering, Macquarie University, Sydney, Australia (e-mail:shihao.yan@mq.edu.au).  J.~Yuan is with the School of Electrical Engineering and Telecommunications, University of New South Wales, Sydney, Australia (e-mail:j.yuan@unsw.edu.au).}
\thanks{We are thankful to L.~S.~Muppirisetty, T.~Charalambous, and H.~Wymeersch with the Department of Signals and Systems, Chalmers University of Technology for useful discussions that helped improve our manuscript.}}
\maketitle

\begin{abstract}
We propose a location-aware pilot allocation algorithm for a massive multiple-input multiple-output (MIMO) network with high-mobility users, where the wireless channels are subject to Rician fading. Pilot allocation in massive MIMO is a hard combinatorial problem and depends on the locations of users. As such, it is highly complex to achieve the optimal pilot allocation in real-time for a network with high-mobility users. Against this background, we propose a low-complexity pilot allocation algorithm, which exploits the behavior of line-of-sight (LOS) interference among the users and allocate the same pilot sequence to the users with small LOS interference. Our examination demonstrates that our proposed algorithm significantly outperforms the existing algorithms, even with localization errors. Specifically, for the system considered in this work, our proposed algorithm provides up to 37.26\% improvement in sum spectral efficiency (SE) and improves the sum SE of the worst interference-affected users by up to 2.57 bits/sec/Hz, as compared to the existing algorithms.
\end{abstract}

\section{Introduction}
Massive multiple-input multiple-output (MIMO) is considered as a key technology for the fifth generation networks, since it offers an increased data rate and improved spectral efficiency (SE). However, challenges such as interference due to
pilot contamination prevent us from achieving the full benefit offered by massive MIMO \cite{Fernandes2013,Jose2011}. Pilot contamination affects the performance of massive MIMO even when the number of antennas at the base station (BS) is very large. Consequently, pilot contamination is a widely studied problem in massive MIMO. The existing studies on pilot contamination can be classified into five broad categories: 1) protocol based \cite{Fernandes2013}; 2) precoding based \cite{Jose2011}; 3) angle-of-arrival (AoA) based \cite{Akbar2016b}; 4) blind \cite{Muller2014}; and 5) pilot design methods \cite{Akbar2016a}.

Recently, an increasing attention has been paid to utilize the location information of users for mitigating pilot contamination. The users' locations can be estimated or requested from the users directly by the BS and then the location information can be leveraged to allocate pilot sequences in the network such that pilot contamination is minimized (e.g.,\cite{Akbar2016b,Muppirisetty2015,Wang2015}). We highlight that the existing location-aware pilot allocation algorithms only consider a simple channel model \cite{Muppirisetty2015,Zhao2015,Wang2015} that may not be generalized enough to depict certain practical channel conditions, such as the channels containing both line-of-sight (LOS) and non-line-of-sight (NLOS) components. In some cases, a LOS channel component may exist between BSs and users \cite{Akbar2016b}. In order to deal with the pilot allocation problem under LOS conditions or when both LOS and NLOS conditions exist, we consider Rician fading channels in this work. Most existing studies in location-aware pilot allocation algorithms assume that the AoAs of all the users are strictly non-overlapping \cite{Muppirisetty2015,Zhao2015,Wang2015}, which is hard to justify in some practical scenarios. We relax this assumption and demonstrate that the location information is beneficial for pilot allocation, even when the AoAs of interfering users are overlapping. Furthermore, different from existing studies, we assume that the pilot sequences used in a cell are not orthogonal. As such, our system model incorporates both the inter-cell and the intra-cell pilot contamination. The work \cite{Akbar2016b} presented the pilot allocation in a single-cell network. Different from \cite{Akbar2016b}, we consider a more general multi-cell network, which encompasses the single-cell network as a special case. Moreover, in this work we derive the multi-cell LOS interference expression that is valid for an arbitrary number of BS antennas. Furthermore, in this work we perform a thorough comparison between the proposed algorithm and several existing algorithms under varying network conditions. We highlight that the derivation and the comparison were not presented in \cite{Akbar2016b}. Throughout this paper, we define the LOS interference as the interference caused by the LOS components in the channels.

In this correspondence, we propose a low-complexity pilot allocation algorithm suitable for a network with high-mobility users. Our examination shows that the proposed algorithm improves the sum SE of the network as compared to the existing algorithms, even when the locations of the users suffer from estimation errors. 

\section{System Model}
\begin{figure}[!t]
\centering
{\includegraphics[width=5.2cm,height=4.6cm]{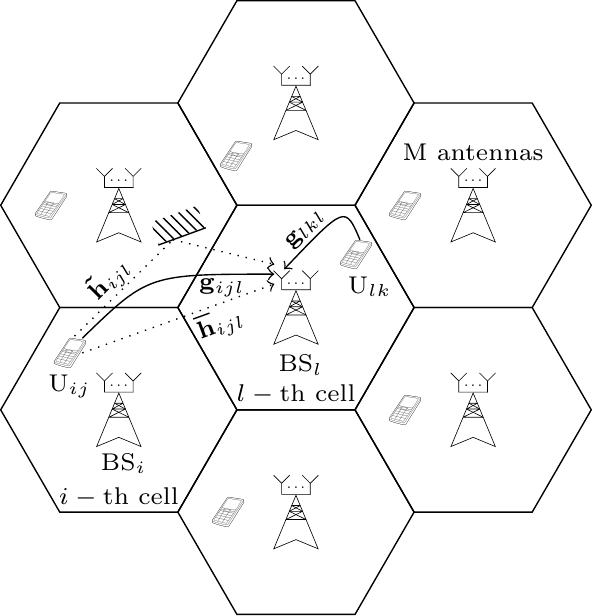}}
\caption{A multi-cell massive MIMO network with Rician fading channels.}
\label{sys_mod}
\end{figure}

We consider an $L$-cell massive MIMO network as illustrated in Fig.~\ref{sys_mod}. In each cell, a BS equipped with $M$ antennas communicates with $N$ single-antenna users. We denote the BS in the \textit{i}-th cell as $\textrm{BS}_i$ and the \textit{j}-th user in the \textit{i}-th cell as $\textrm{U}_{ij}$. Additionally, we denote the location of $\textrm{U}_{ij}$ as $(d_{ij},\theta_{ij})$, where $d_{ij}$ is the distance from $\textrm{U}_{ij}$ to $\textrm{BS}_i$ and $\theta_{ij}$ is the AoA of $\textrm{U}_{ij}$ at $\textrm{BS}_i$. We represent the small-scale propagation factor between $\textrm{U}_{ij}$ and the $\textrm{BS}_l$ as $\mathbf{h}_{ijl}$. We assume that $\mathbf{h}_{ijl}$ is subject to Rician fading. Consequently, $\mathbf{h}_{ijl}$ consists of a LOS component denoted as $\mathbf{\bar{h}}_{ijl}$ and a Rayleigh distributed NLOS component denoted as $\tilde{\mathbf{h}}_{ijl}$, where $\tilde{\mathbf{h}}_{ijl} \sim \mathcal{CN}(\mathbf{0},\mathbf{I}_M)$. We assume that the uplink channel between $\textrm{U}_{ij}$ and $\textrm{BS}_l$ is affected by large-scale propagation effects denoted as $\alpha_{ijl}$. As such, the uplink channel from $\textrm{U}_{ij}$ to $\textrm{BS}_l$ is written as
\small
\begin{align}\label{chan_eq}
\mathbf{g}_{ijl} &= \sqrt{\alpha_{ijl}}\left(\sqrt{\frac{K_{ijl}}{1+K_{ijl}}}\mathbf{\bar{h}}_{ijl} + \sqrt{\frac{1}{1+K_{ijl}}}\tilde{\mathbf{h}}_{ijl} \right),
\end{align}
\normalsize
where $K_{ijl}$ is the \textit{K}-factor of $\textrm{U}_{ij}$ at $\textrm{BS}_{l}$. We assume that the BSs are equipped with uniform linear antenna arrays. As such, $\mathbf{\bar{h}}_{ijl}$ depends on the location of $\textrm{U}_{ij}$ and is expressed as
\begin{align}\label{LOS_def}
  \mathbf{\bar{h}}_{ijl} &=[1,e^{-j\left(\frac{2\pi r}{\lambda}\right)\textrm{sin}\left(\theta_{ij}\right)},\dotsc,e^{-j\left(M-1\right)\left(\frac{2\pi r}{\lambda}\right)\textrm{sin}\left(\theta_{ij}\right)}]^T,
\end{align}
where $r$ is the distance between two antennas and $\lambda$ is the wavelength. We assume that $r=\frac{\lambda}{2}$, which is a widely adopted assumption \cite{Akbar2016b}. The uplink channels between all the users in the \textit{i}-th cell and $\textrm{BS}_l$ are represented as $\mathbf{G}_{il} = \mathbf{H}_{il}\mathbf{D}_{il}^{\frac{1}{2}}$, where
\begin{align}\label{uplink_chan}
  \mathbf{H}_{il} &= \mathbf{\bar{H}}_{il}[\mathbf{\Omega}_{il}\left(\mathbf{\Omega}_{il}+\mathbf{I}_N\right)^{-1}]^{\frac{1}{2}} + \tilde{\mathbf{H}}_{il}[\left(\mathbf{\Omega}_{il}+\mathbf{I}_N\right)^{-1}]^{\frac{1}{2}},
\end{align}
with $\mathbf{\bar{H}}_{il}=[\mathbf{\bar{h}}_{i1l},\dotsc,\mathbf{\bar{h}}_{iNl}]$, $\tilde{\mathbf{H}}_{il}=[\tilde{\mathbf{h}}_{i1l},\dotsc,\tilde{\mathbf{h}}_{iNl}]$, $\mathbf{\Omega}_{il}=\textrm{diag}[K_{i1l},\dotsc,K_{iNl}]$, and $\mathbf{D}_{il} = \textrm{diag}[\alpha_{i1l},\dotsc,\alpha_{iNl}]$.

In this work, we consider the uplink transmission from the users to the BSs, which consists of two phases, i.e., uplink channel estimation and uplink data transmission.

\subsection{Uplink Channel Estimation}
In the uplink channel estimation phase, users from all the cells transmit their pre-assigned pilot sequences to the BSs. We assume that the length of the pilot sequence is $\ell$. As such, only $\ell$ orthogonal pilot sequences are available in the massive MIMO network. The pilot sequence assigned to $\textrm{U}_{ij}$ is represented as $\bm{\eta}_{ij}$. Accordingly, the pilot sequences assigned to all the users in the \textit{i}-th cell are represented as
$\mathbf{\Lambda}_i=[\bm{\eta}_{i1},\bm{\eta}_{i2},\dotsc,\bm{\eta}_{iN}]^T$. We assume that the pilot sequences are transmitted with unit power. The received matrix at $\textrm{BS}_l$ in the uplink pilot transmission phase is given by
\begin{align}\label{uplink_pilot}
\mathbf{Y}_l &= \mathbf{G}_{ll}\mathbf{\Lambda}_l + \textrm{$\textstyle \sum_{i=1,i\neq l}^L$}\mathbf{G}_{il}\mathbf{\Lambda}_i + \mathbf{Z},
\end{align}
where $\mathbf{Z}$ is the Gaussian noise matrix at the $\textrm{BS}_l$ and the distribution of each independent element in $\mathbf{Z}$ follows $ \mathcal{CN}(0,\sigma_z^2)$. We assume that $\textrm{BS}_l$ only knows the estimated location of each user, i.e, $(\hat{d}_{ij},\hat{\theta}_{ij})$. As such, the estimates of the LOS components in \eqref{uplink_pilot} are known at $\textrm{BS}_l$. Next, we compute the received matrix corresponding to the NLOS component by subtracting the LOS components from $\mathbf{Y}_l$, i.e., $\hat{\tilde{\mathbf{Y}}}_l = \mathbf{Y}_l-\hat{\bar{\mathbf{Y}}}_l$, where $\hat{\bar{\mathbf{Y}}}_l = \sum_{i=1}^L \mathbf{\hat{\bar{H}}}_{il}\mathbf{\hat{D}}_{il}^\frac{1}{2}[\mathbf{\hat{\Omega}}_{il}(\mathbf{\hat{\Omega}}_{il}+ \mathbf{I}_N)^{-1}]^{\frac{1}{2}}\mathbf{\Lambda}_i$, $\mathbf{\hat{D}}_{il}=\textrm{diag}[\hat{\alpha}_{i1l},\dotsc,\hat{\alpha}_{iNl}]$, $\mathbf{\hat{\bar{H}}}_{il}=[\mathbf{\hat{\bar{h}}}_{i1l},\dotsc,\mathbf{\hat{\bar{h}}}_{iNl}]$, $\mathbf{\hat{\Omega}}_{il}=\textrm{diag}[\hat{K}_{i1l},\dotsc,\hat{K}_{iNl}]$. We assume that $\mathbf{\hat{\bar{h}}}_{ijl}$ is obtained from \eqref{LOS_def} using $\hat{\theta}_{ij}$ and we obtain $\hat{\alpha}_{i1l}$ and $\hat{K}_{i1l}$ based on the estimated distance $\hat{d}_{ij}$. Accordingly, $\hat{\tilde{\mathbf{Y}}}_l$ is obtained from \eqref{uplink_pilot} as
\begin{align}\label{uplink_pilot_nlos}
&\hat{\tilde{\mathbf{Y}}}_l =\tilde{\mathbf{H}}_{ll}[\left(\mathbf{\Omega}_{ll}+\mathbf{I}_N\right)^{-1}]^{\frac{1}{2}} \mathbf{D}_{ll}^\frac{1}{2} \mathbf{\Lambda}_l  \notag \\ &+  \textrm{$\textstyle \sum_{i=1,i\neq l}^L$}\tilde{\mathbf{H}}_{il}[\left(\mathbf{\Omega}_{il}+\mathbf{I}_N\right)^{-1}]^{\frac{1}{2}}\mathbf{D}_{il}^\frac{1}{2} \mathbf{\Lambda}_i+ \textrm{$\textstyle \sum_{i=1}^L$}\mathbf{\xi}_{il} + \mathbf{Z},
\end{align}
where
$\xi_{il} = \bar{\mathbf{H}}_{il}\mathbf{D}_{il}^\frac{1}{2}[\mathbf{{\Omega}}_{il}(\mathbf{{\Omega}}_{il}+\mathbf{I}_N)^{-1}]^{\frac{1}{2}}\mathbf{\Lambda}_i- \hat{\bar{\mathbf{H}}}_{il}\mathbf{\hat{D}}_{il}^\frac{1}{2}[\mathbf{\hat{\Omega}}_{il}(\mathbf{\hat{\Omega}}_{il}+\mathbf{I}_N)^{-1}]^{\frac{1}{2}} \mathbf{\Lambda}_i$. We highlight that $\xi_{il}$ appears in \eqref{uplink_pilot_nlos} due to localization errors. If the locations of users are precisely known, we have $\xi_{il}=0$ and the received matrix corresponding to the LOS component is completely removed from \eqref{uplink_pilot}. In other words, without localization error the AoA of each user does not affect the channel estimation.

We assume that the $\textrm{BS}_l$ obtains the least-square (LS) channel estimates from \eqref{uplink_pilot_nlos} as
\begin{align}\label{ls_estimate}
\hat{\tilde{\mathbf{G}}}_{ll}= \tilde{\mathbf{Y}}_l\mathbf{\Lambda}_l^{H}&=\tilde{\mathbf{H}}_{ll}[\left(\mathbf{\Omega}_{ll}+\mathbf{I}_N\right)^{-1}]^{\frac{1}{2}} \mathbf{D}_{ll}^\frac{1}{2}\mathbf{R}_{ll} \notag \\ &+ \textrm{$\textstyle \sum_{\substack{i=1 \\ i\neq l}}^L$}\tilde{\mathbf{H}}_{il} [\left(\mathbf{\Omega}_{il}+\mathbf{I}_N\right)^{-1}]^{\frac{1}{2}}\mathbf{D}_{il}^\frac{1}{2}\mathbf{R}_{il} + \bar{\mathbf{Z}},
\end{align}
where $\mathbf{R}_{il}=\mathbf{\Lambda}_i\mathbf{\Lambda}_l^{H}$, and $\bar{\mathbf{Z}}= \textrm{$\textstyle \sum_{i=1}^L$}\mathbf{\xi}_{il}\mathbf{\Lambda}_l^{H} +\mathbf{Z\Lambda}_l^{H}$.
\begin{remark}
From \eqref{ls_estimate}, we note that the channel estimate $\hat{\tilde{\mathbf{G}}}_{ll}$ suffers from intra-cell pilot contamination when $\mathbf{R}_{ll}\neq\mathbf{I}_N$. Additionally, we note that the channel estimates suffer from inter-cell pilot contamination when the same pilot sequences are repeated throughout the network. Specifically, when $\mathbf{R}_{il} \neq \mathbf{0}_N$ the channel estimates suffer from inter-cell pilot contamination. In real-world telecommunication networks, it is not possible to assign orthogonal pilot sequences to all the users in the network. As such, it is reasonable to assume that $\mathbf{R}_{il} \neq \mathbf{0}_N$ and inter-cell pilot contamination always exists.
\end{remark}

\subsection{Uplink Data Transmission and Spectral Efficiency}
During the uplink data transmission, each user in a cell transmits uplink data symbols to the same-cell BSs. The $\textrm{BS}_l$ multiplies the received vector $\mathbf{y}_l$ with the zero-forcing (ZF) matrix $\mathbf{W}_l$ to decode the symbols transmitted by the same-cell users. Accordingly, using the use-and-then-forget bound, the signal received from $\textrm{U}_{lk}$ after detection is written as \cite{Emil2016}
\begin{align}\label{received_ulk}
\mathbf{w}_{lk}^{H}\mathbf{y}_{l}  &= \mathbb{E}\{\mathbf{w}_{lk}^H\mathbf{g}_{lkl}\}x_{lk} +(\mathbf{w}_{lk}^H\mathbf{g}_{lkl} - \mathbb{E}\{\mathbf{w}_{lk}^H\mathbf{g}_{lkl} \})x_{lk} \notag \\
&+ \textrm{$\textstyle \sum_{i=1}^L \sum_{j=1, (i,j)\neq (l,k)}^N$} \mathbf{w}_{lk}^H \mathbf{g}_{ijl}x_{ij} + \nicefrac{\mathbf{w}_{lk}^H \mathbf{n}_l}{\sqrt{\rho}},
\end{align}
where $\mathbf{w}_{lk}$ is the \textit{k}-th column of the ZF detection matrix $\mathbf{W}_l$, $x_{ij}$ is the symbol transmitted by $\textrm{U}_{ij}$, $\mathbf{n}_{l}$ is the Gaussian noise at $\textrm{BS}_l$, and $\rho$ is the signal-to-noise ratio. The uplink SE for $\textrm{U}_{lk}$ is given as $\textrm{SE}_{lk}=\left(1-\nicefrac{\ell}{\ell_c}\right)\log_2\left(1+\varrho_{lk}\right)$, where $\ell_c$ is the channel uses in a coherence block \cite{Emil2016}, and $\varrho_{lk}$ is the signal-to-interference-plus-noise ratio (SINR) given by
\small
\begin{align}\label{sinr_ulk}
   \varrho_{lk} = \frac{|\mathbb{E}[{\mathbf{w}}_{lk}^H{\mathbf{g}}_{lkl}]|^2}{\sum\limits_{i=1}^L\sum\limits_{j=1}^N \mathbb{E}[|{\mathbf{w}}_{lk}^H{\mathbf{g}}_{ijl}|^2] - |\mathbb{E}[{\mathbf{w}}_{lk}^H{\mathbf{g}}_{lkl}]|^2 + \frac{1}{\rho}\mathbb{E}\{\|\mathbf{w}_{lk}\|^2\}}.
\end{align}
\normalsize

We note that the product of linear detection vector and the channel, i.e., ${\mathbf{w}}_{lk}^H{\mathbf{g}}_{ijl}$, is important in determining $\varrho_{lk}$. We next present the proposed pilot allocation algorithm that aims at reducing the interference caused by ${\mathbf{w}}_{lk}^H{\mathbf{g}}_{ijl}$ based on the estimated locations of the users.

\section{Location-Aware Pilot Allocation}
In this section, we present our low-complexity location-aware pilot allocation algorithm. The algorithm requires the knowledge about the large-scale fading, AoAs, and $K$-factors to perform pilot allocation. Specifically, we first derive the expression for the LOS interference based on the estimated locations of the users. Then, the pilot sequences are allocated to the users sequentially to minimize the LOS interference.
\subsection{LOS Interference}
\begin{figure}
\centering
\includegraphics[width=8.0cm,height=4.8cm]{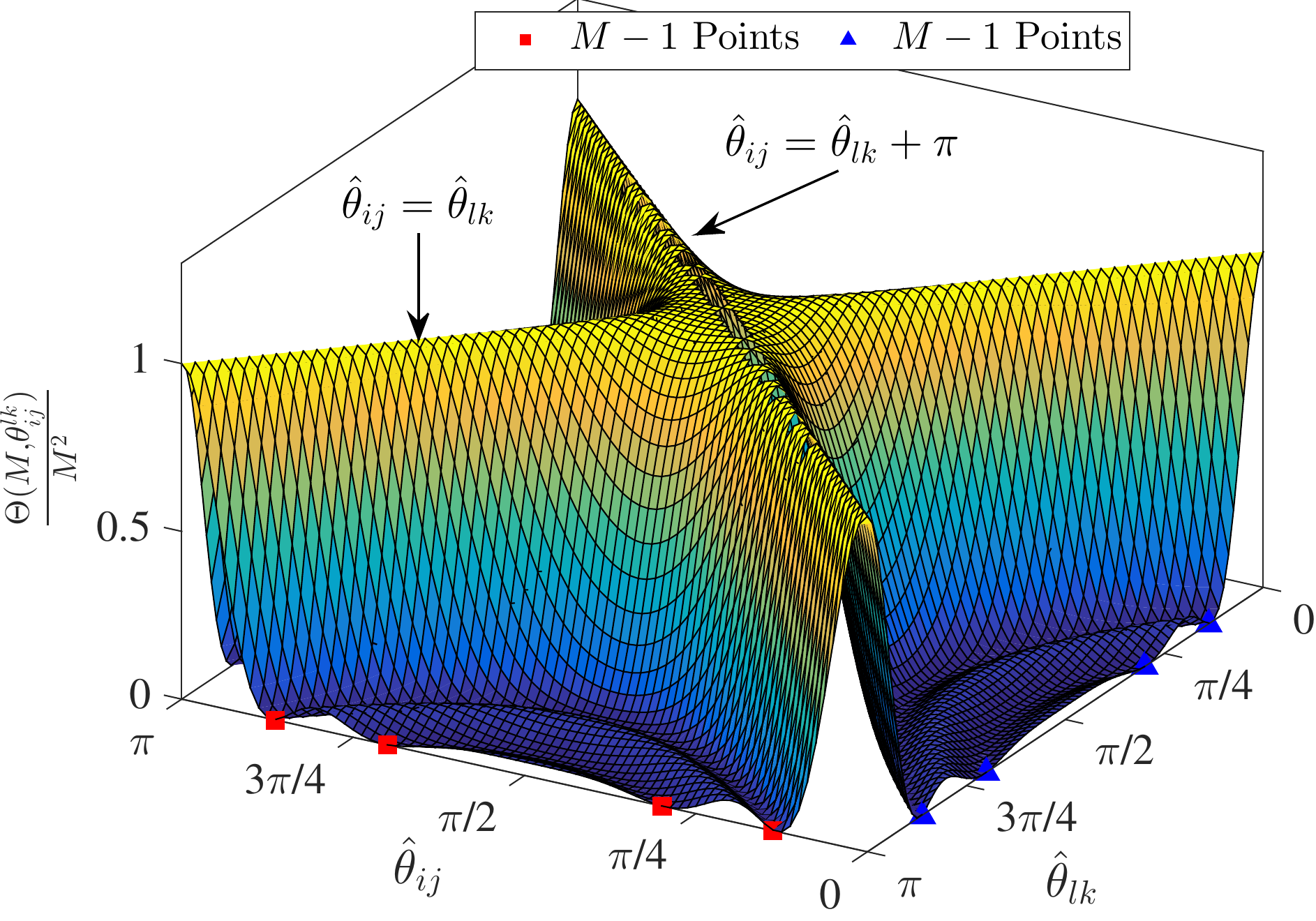}
\caption{AoA-dependent part of the LOS interference versus the AoAs.}
\label{capacity}
\end{figure}
We derive the expression for the LOS interference between two users in the network in the following theorem.
\begin{theorem}\label{los_interference}
The LOS interference between $\textrm{U}_{ij}$ and $\textrm{U}_{lk}$ based on their estimated locations is given as
\begin{align}\label{interference_function}
   \delta_{ij}^{lk} &= \underbrace{\frac{\hat{\alpha}_{ijl}K_{ijl}\left(1+K_{lkl}\right)} {\hat{\alpha}_{lkl}K_{lkl}\left(1+K_{ijl}\right)}}_{\textrm{part 1}}\underbrace{\frac{|\Theta(M,\hat{\theta}_{ij}^{lk})|^2}{M^2}}_{\textrm{part 2}},
\end{align}
where
\small
\begin{align}\label{theta_def}
\hspace{-0.34cm} \left|\Theta(M,\hat{\theta}_{ij}^{lk})\right|^2 \hspace{-0.15cm} &= \hspace{-0.10cm} \begin{cases}
      M^2 & \hspace{-0.15cm}  \hat{\theta}_{ij}^{lk} = 0, \\
     \frac{\sin^2\left(M\hat{\theta}_{ij}^{lk}/2\right)}{\sin^2\left(\hat{\theta}_{ij}^{lk}/2\right)} & \hspace{-0.15cm} -2\pi < \hat{\theta}_{ij}^{lk} < 2\pi~\textrm{and}~\hat{\theta}_{ij}^{lk}\neq 0.
\end{cases}
\end{align}
\normalsize
\end{theorem}
\begin{IEEEproof}
The proof is provided in Appendix \ref{int_appendix}.
\end{IEEEproof}

We note that the LOS interference given in \eqref{interference_function} consists of two parts. Specifically, \textit{part 1} is distance-dependent and \textit{part 2} is AoA-dependent. We note that \textit{part 1} is the smallest when the $\textrm{U}_{ij}$ and $\textrm{U}_{lk}$ are  the farthest apart from each other. Fig.~\ref{capacity} depicts \textit{part 2} at different AoAs for $\textrm{U}_{ij}$ and $\textrm{U}_{lk}$ for $M=5$. We obtain the following insights regarding AoA-dependent \textit{part 2} from Fig.~\ref{capacity}. The AoA-dependent \textit{part 2} is
\begin{itemize}
\item maximum when the AoAs are overlapping, i.e., $\hat{\theta}_{ij}=\hat{\theta}_{lk}$.
\item maximum when the AoAs differ by $\pi$, i.e., $\hat{\theta}_{ij}=\pi-\hat{\theta}_{lk}$.
  \item minimum for certain mutual AoAs for $\textrm{U}_{ij}$ and $\textrm{U}_{lk}$.
  \item minimum for $M-1$ possible values of $\hat{\theta}_{ij}$ or $\hat{\theta}_{lk}$.
\end{itemize}

We highlight that the observations on \textit{part 1} and \textit{part 2} can be utilized for pilot allocation. As such, the location information can be used to identify the users with the minimum LOS interference and assign the same pilot sequence to such users to improve the sum SE. We note that the LOS interference \eqref{interference_function} is obtained for the LS estimator and the ZF detector. We also clarify that for different combination of estimators and detectors, the LOS interference exists and can be obtained in a similar way to obtaining \eqref{interference_function}.
\begin{lemma}\label{lemma1}
In massive MIMO networks when $M\rightarrow \infty$, we have $\delta_{ij}^{lk} \xrightarrow{a.s} 0$ when $\hat{\theta}_{ij} \neq \hat{\theta}_{lk}$ and
\begin{align}\label{mmimo}
  \delta_{ij}^{lk} \xrightarrow{a.s} \frac{\hat{\alpha}_{ijl}K_{ijl}\left(1+K_{lkl}\right)} {\hat{\alpha}_{lkl}K_{lkl}\left(1+K_{ijl}\right)}~\textrm{when $\hat{\theta}_{ij} = \hat{\theta}_{lk}$}.
\end{align}
\end{lemma}
\begin{IEEEproof}
The proof follows directly from the definition of $|\Theta(M,\hat{\theta}_{li}^{jk})|^2$ in \eqref{theta_def}. When $\hat{\theta}_{ij}^{lk}\neq 0$, the minimum value of \eqref{theta_def} is obtained, i.e., $|\Theta(M,\hat{\theta}_{li}^{jk})|^2=0$, when $\hat{\theta}_{ij}^{lk}=2b\pi/M$ with $ b=\pm 1,\dotsc,\pm (M-1)$. Moreover, when $M\rightarrow \infty$, we find that $|\Theta(M,\hat{\theta}_{ij}^{lk})|^2\xrightarrow{a.s}0$ for $\hat{\theta}_{ij}^{lk}\neq 0$. Substituting \eqref{theta_def} into \eqref{interference_function}, we obtain \eqref{mmimo}, which completes the proof.
\end{IEEEproof}
\begin{remark}
According to \textit{Lemma}~\ref{lemma1}, when the number of antennas at BSs is large, the LOS interference is high only when the AoAs of the two interfering users are strictly overlapping. Additionally, increasing the number of antennas at BSs increases the possibility of having two users with the minimum LOS interference, because the LOS interference is minimum at $M-1$ values of the mutual AoAs, as depicted in Fig.~\ref{capacity}. This observation highlights the benefit of massive MIMO for the proposed location-aware pilot allocation algorithm.
\end{remark}
\begin{remark}
\noindent The optimal solution for pilot allocation can be found by performing an exhaustive search to identify the pilot allocation that leads to the highest sum SE. However, this exhaustive search is of a high computational complexity, which makes it infeasible for the network with a large number of high-mobility users. For example, for a given $N$ and $\ell$, there are $L\times {N^\ell}$ possible pilot allocations to be searched in each cell. This motivates us to propose a low-complexity pilot allocation algorithm in the next subsection.
\end{remark}
\subsection{Pilot Allocation Algorithm}
In this subsection, we detail the proposed low-complexity pilot sequence allocation algorithm, which results in reduced pilot contamination and improved sum SE. We note that the SINR expression in \eqref{sinr_ulk} depends on instantaneous channel realizations, which cannot be accurately obtained when the network suffers from pilot contamination \cite{Zhu2015}. Due to this limitation, we focus on minimizing the LOS interference given by \eqref{interference_function} and allocate the same pilot sequence to the users with low LOS interferences.

We next present the step-by-step algorithm for pilot sequence allocation. We first assign pilot sequences to the center cell and then to the neighboring cells.

\subsubsection{Divide the cell in to tiers}
The BS divides the cell area into $n=\textrm{ceil}(\nicefrac{N}{\ell})$ tiers based on the estimated distance $\hat{d}_{lk}$ as
\small
\begin{align}\label{tiers}
[\underbrace{\hat{d}_{l1},\dotsc,\hat{d}_{l\ell}}_{\textrm{tier 1}} \underbrace{\hat{d}_{l(\ell+1)},\dotsc,\hat{d}_{l(2\ell)}}_{\textrm{tier 2}},\dotsc,\underbrace{\hat{d}_{l\{(n-1)\ell+1\}},\dotsc,\hat{d}_{lN}}_{\textrm{tier $n$}}],
\end{align}
\normalsize
where each of the first $n-1$ tiers consists of $\ell$ users while there are no more than $\ell$ users in the tier $n$.
\subsubsection{Assign pilots in tier 1}
We next assign the orthogonal pilot sequences to the $\ell$ users in the tier 1, i.e., the tier closest to the BS. The rationale behind assigning orthogonal pilot sequences to the users in tier 1 is to reduce the pilot contamination. Specifically, the interference power from the users that are assigned the same pilot sequence depends on the large-scale channel coefficient, i.e, $\alpha_{ijl}$. As such, if two users close to a BS are assigned the same pilot sequence, the interference is large, which results in an increased pilot contamination. This observation can also be validated from the distance-dependent \textit{part 1} in \eqref{interference_function}.

\subsubsection{Assign pilots in tier 2}
We compute $\delta_{ij}^{ik}$ using \eqref{interference_function} between $\textrm{U}_{ij}$ in the tier 1 and the $\ell$ users in the tier 2. We highlight that we first assign pilot to the user closest to the BS. We assign the same pilot sequence as $\textrm{U}_{ij}$ in tier 1 to the user $\textrm{U}_{ik}$ in tier 2 with minimum $\delta_{ij}^{ik}$. If $K_{ijl}$ or $K_{lkl}$ are zero, we only used \textit{part 2} in \eqref{interference_function}. We repeat this process for the $\ell-1$ remaining users in tier 2.

\subsubsection{Assign pilots in remaining tiers}
We compute the LOS interference $\delta_{ij}^{ik}$ between $\textrm{U}_{ik}$ in tier $n$ and $\textrm{U}_{ij}$ in tier $o$, where $o\in\{1,\dotsc,n-1\}$ and $\textrm{U}_{ij}$ in tier $o$ has been assigned the same pilot sequence. We then compute the average of the LOS interference and assign the same pilot sequence to $\textrm{U}_{ik}$ with the minimum average LOS interference. If the tier $n$ has less than $\ell$ users, some pilot sequences are not used in tier $n$.

\subsubsection{Assign pilots in remaining cells}
After completing pilot allocation for the center cell, we repeat Step 1 to Step 4 for pilot sequence allocation in neighboring cells. In Step 5, we consider all the LOS interference between a tier in the target cell and all the tiers in cells where the pilot sequence allocation has already been carried out, and then compute the average LOS interference and assign pilot sequences accordingly.

The algorithm returns a vector $\mathbf{s}$, where the \textit{l}-th element of $\mathbf{s}$ denotes the pilot allocation for the users in the \textit{l}-th cell.

\section{Numerical Results and Analysis}
In this section, we demonstrate the benefits of the proposed pilot allocation algorithm with random pilot allocation \cite{Yin2013}, the greedy iterative algorithm \cite{Ngo2017}, and sector-based \cite{Wang2016} as benchmarks. Random pilot allocation is the most widely adopted pilot allocation algorithm in massive MIMO \cite{Muppirisetty2015,Akbar2016a,Akbar2016b}. In random pilot allocation, the BS allocates the pilots to all the users in a cell randomly. Greedy iterative algorithm \cite{Ngo2017} iteratively refines the sum rate by first identifying the user with the lowest rate and then searching a pilot sequence for the user which minimizes the interference. Sector-based algorithm divides the cell-area in sectors and all users in a sector are assigned the same pilot \cite{Wang2016}. The system settings adopted in this section are summarized in Table~\ref{sim_par}. All the results are obtained for an average of 10,000 Monte Carlo simulations.

\begin{figure*}[!t]
\centering
\begin{minipage}{\textwidth}
\subfloat[]{\includegraphics[width=6.0cm,height=5.0cm]{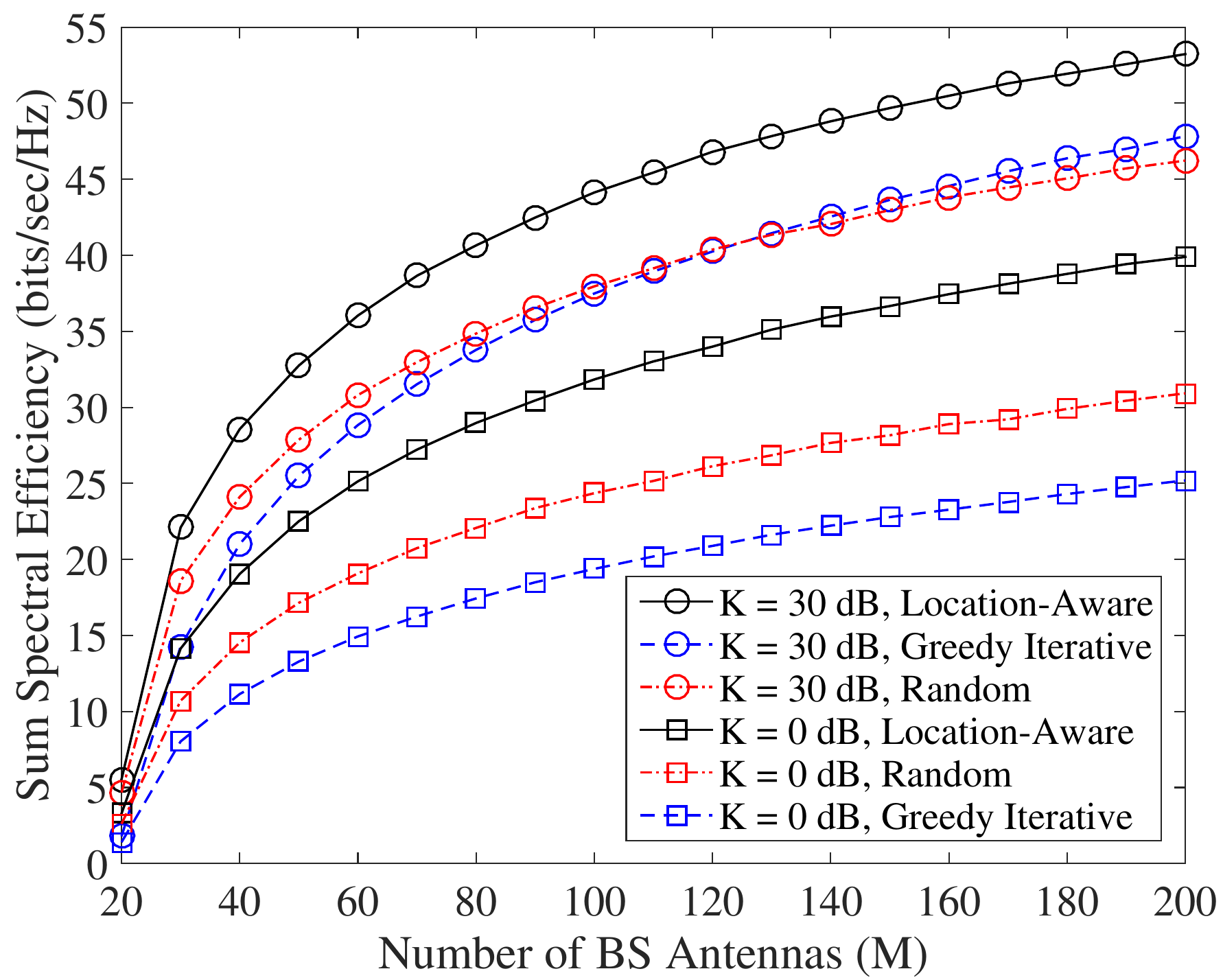} \label{figa}}
\subfloat[]{\includegraphics[width=6.0cm,height=5.0cm]{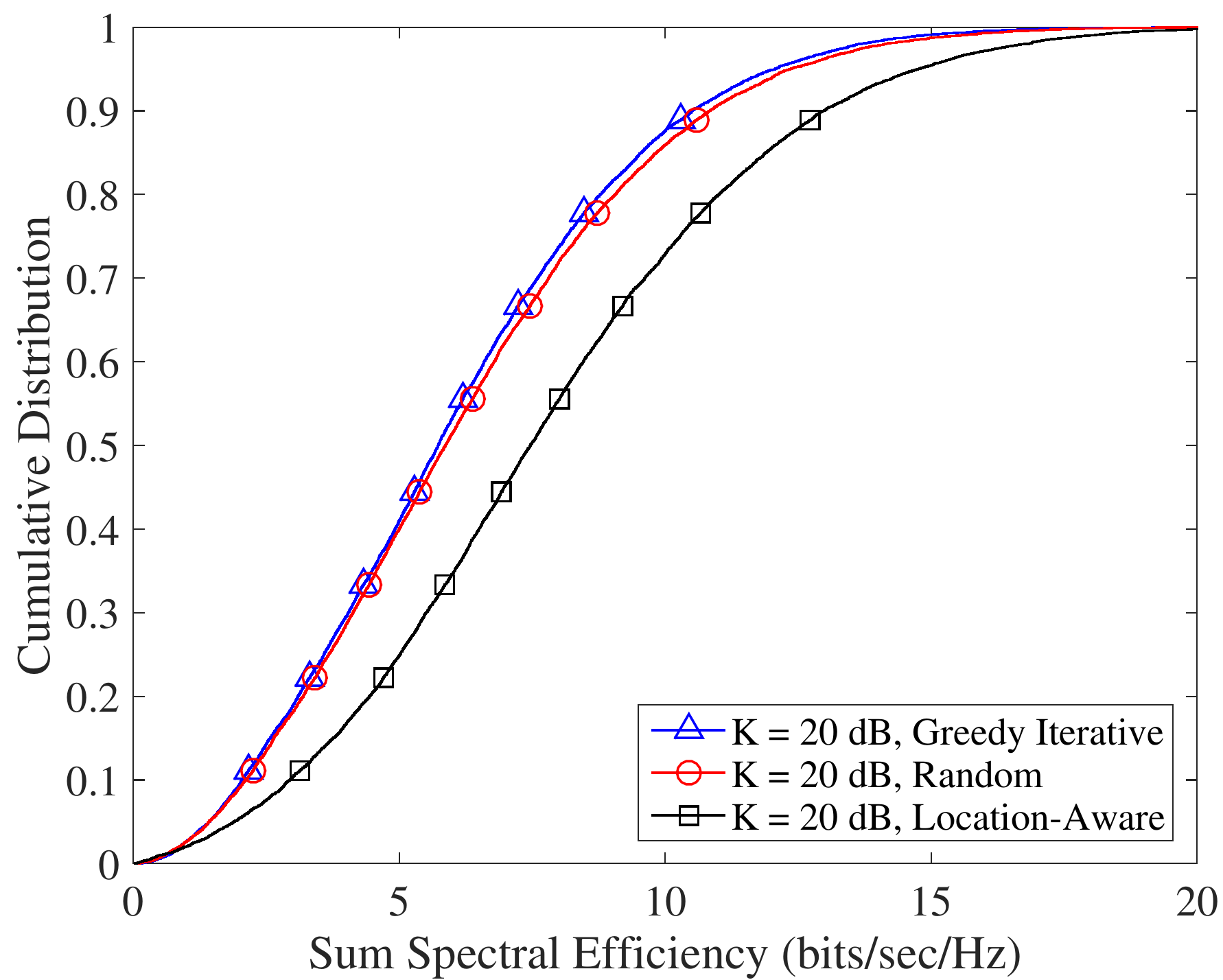} \label{figb}}
\subfloat[]{\includegraphics[width=6.0cm,height=5.0cm]{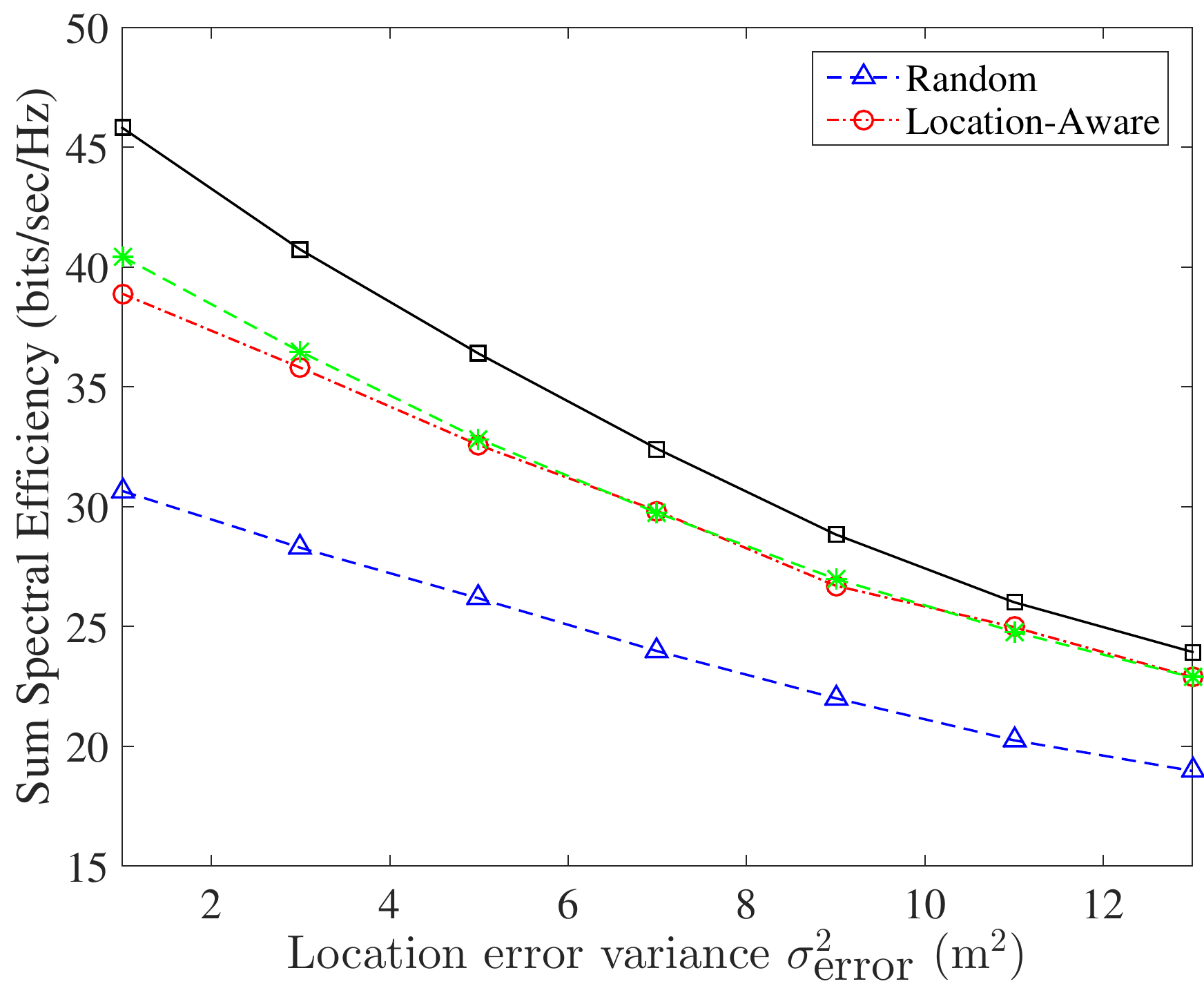} \label{figc}}
\caption{Numerical results illustrating the advantages of the proposed location-aware algorithm. (a) The sum SE versus the number of BS antennas, (b) The CDF of the sum SE for $M=100$, (c) The sum SE versus location error variance for the proposed location-aware algorithm and the three existing algorithms.}
\end{minipage}
\end{figure*}

Fig.~3\subref{figa} depicts the sum SE in the center cell for the proposed pilot allocation, random pilot allocation, and greedy iterative algorithm\footnote{We note that there are $L\times N^{\ell}$ unique pilot allocations in the massive MIMO network, which leads to that identifying the optimal pilot allocation is of a high computational complexity. For a small scale scenario with $N=4$, $\ell=2$, and $L=1$, we have found that the proposed pilot allocation algorithm achieves between 67.7\% to 76.6\% of the sum SE achieved by the optimal pilot allocation algorithm, but with a significantly lower computational complexity.}. We assume that the $K$ factors for all the users are the same. The results are obtained using \eqref{sinr_ulk} for an average of 10,000 random user locations, where the user locations are accurately known to the BSs. By randomizing user locations we simulate high-mobility scenarios, where the locations of users change for each coherence interval. The advantage of the proposed pilot allocation algorithm is clearly observed from Fig.~3\subref{figa}, where for $M=200$ and $K=0\,\textrm{dB}$ the proposed pilot allocation algorithm provides $22.73\%$ and $37.26\%$ improvement in sum SE as compared to the random pilot allocation and greedy iterative approach, respectively. Importantly, we observe that the performance improvement increases with $K$, which is due to the fact that the LOS interference dominates the total interference when $K$ is large and our proposed algorithm is to minimize the LOS interference. We also observe that the performance improvement increases with $M$, which can be explained by our \textit{Lemma 1} and demonstrates the benefit of massive MIMO. We observe that the greedy iterative algorithm achieves a lower sum SE than the random pilot allocation algorithm when $K$ is small. This is due to the fact that in high-mobility scenarios, the user with low SINR may have a different location and channel conditions in the next iteration. This makes the greedy iterative algorithm more suitable for low-mobility scenarios.

\begin{table}[!t]
\renewcommand{\arraystretch}{1.0}
\caption{Simulation Parameters}
\label{sim_par}
\centering
\scalebox{0.78}{\begin{tabular}{|l|l|}
\hline \hline
Number of cells $L$ & 2  \\\hline
Number of users per cell $N$ & 36 \\ \hline
Cell radius $d_{\textrm{BS}}$ & 400\,m \\ \hline
User distance from the BS $d_{lk}$ & $U\left[100\,\textrm{m},400\,\textrm{m}\right]$ \\ \hline
User AoA $\theta_{lk}$ & $U\left[0,2\Lambda\right]$ \\ \hline
Large-scale propagation constant $\alpha_{lk}$ & $(\nicefrac{\hat{d}_{lk}}{d_{\textrm{BS}}})^v$, where $v=3.76$\\ \hline
Length of pilot sequence $\ell$ & 12 \\ \hline
Channel uses in coherence interval $\ell_c$ & 196 \\ \hline
SNR $\rho$ & 10\,dB \\
\hline \hline
\end{tabular}}
\end{table}

In Fig.~3\subref{figb}, we compare the performance of the users worst affected by interference. As such, the sum SE of such users is low. We compare the cumulative distribution function (CDF) of five users with minimum sum SE for the three pilot allocation algorithms. We set $M=100$, $K=10\;\textrm{dB}$, and assume that the locations of users are accurately known to the BSs. Our proposed algorithm outperforms the random pilot allocation and greedy iterative algorithm by $1.56\,\textrm{bit/sec/Hz}$ and $2.57\,\textrm{bit/sec/Hz}$, respectively.

In Fig.~3\subref{figc}, we examine the performance of the proposed algorithm subject to localization errors. The location estimate has an error variance of $\sigma_{\textrm{error}}^2$, where the location error is uniformly distributed. We clarify that certain location-based schemes may be more sensitive to the errors in AoA than to the errors in distance. However, in this simulation, we consider the errors in both distance and AoA. In the simulations, we use the $K$-factor as defined in 3GPP TR25.996 model. Accordingly, we define $K_{ijl}~(\textrm{dB})=13-0.03\hat{d}_{ijl}$, where $\hat{d}_{ijl}$ is the estimated distance between $\textrm{U}_{ij}$ and $\textrm{BS}_l$. Furthermore, we assume that the probability of LOS decreases linearly as the distance between users and BSs increases. We highlight that the proposed algorithm still significantly outperforms the other two pilot allocation algorithms when the locations of the users suffer from estimation errors. We note that the sum SE decreases when $\sigma_{\textrm{error}}^2$ increases. For example, when $\sigma_{\textrm{error}}^2$ increases from $0\,\textrm{m}^2$ to $15\,\textrm{m}^2$, the sum SE for the proposed algorithm decreases from $28.08\,\textrm{bit/sec/Hz}$ to $24.59\,\textrm{bit/sec/Hz}$, which amounts to a reduction of $12.43\%$ in the sum SE. However, the proposed pilot allocation algorithm provides an improvement of $10.46\%$, $12.10\%$, and $30.55\%$ in the sum SE, respectively, as compared with sector-based allocation, random allocation, and greedy allocation when $\sigma_{\textrm{error}}^2=3\,\textrm{m}^2$.
\vspace{-0.2in}
\section{Conclusion}
We proposed a low-complexity location-aware pilot allocation algorithm for a massive MIMO network with high-mobility users. The algorithm exploited the behavior of LOS interference between users for pilot sequence allocation. Comparison with existing algorithms
demonstrated the advantages of our proposed algorithm in terms of achieving a higher sum SE. In addition, the proposed algorithm is beneficial for the users that are worst-affected by interference and outperforms existing algorithms in the presence of localization errors.

\appendices
\vspace{-0.2in}
\section{} \label{int_appendix}
From \eqref{sinr_ulk}, we note that ${\mathbf{w}}_{lk}^H{\mathbf{g}}_{ijl}$ is important in determining $\varrho_{lk}$. The vector $\mathbf{w}_{lk}^H$ is based on channel estimates. For example, we have $\mathbf{w}_{lk}=\nicefrac{\hat{\mathbf{g}}_{lkl}}{\hat{\mathbf{g}}_{lkl}^H\hat{\mathbf{g}}_{lkl}}$ for ZF detection. Consequently, we compute $\mathbf{\hat{g}}_{lkl}^H\mathbf{g}_{ijl}$ and obtain
\small
\begin{align}\label{app2}
   \mathbf{\hat{g}}_{lkl}^H\mathbf{g}_{ijl}
   &= \hat{\bar{\mathbf{g}}}_{lkl}^H\mathbf{\bar{g}}_{ijl} + \hat{\bar{\mathbf{g}}}_{lkl}^H\mathbf{\tilde{g}}_{ijl} + \mathbf{\hat{\tilde{g}}}_{lkl}^H\mathbf{\bar{g}}_{ijl} + \mathbf{\hat{\tilde{g}}}_{lkl}^H\mathbf{\tilde{g}}_{ijl}.
\end{align}
\normalsize
We highlight that \eqref{app2} cannot be computed unless the locations of users and channel estimates are known to the BS. However, assuming that the BSs have estimated user locations $(\hat{d}_{ij},\hat{\theta}_{ij})$, we calculate an estimate for the first term, i.e, the pure LOS term in \eqref{app2}, as
\small
\begin{align}\label{app3}
  \mathbf{\hat{\bar{g}}}_{lkl}^H\mathbf{\hat{\bar{g}}}_{ijl} &=  \left(\frac{\hat{\alpha}_{lkl}\hat{\alpha}_{ijl}K_{lkl}K_{ijl}}{\left(1+K_{lkl}\right)\left(1+K_{ijl}\right)}\right)^{\frac{1}{2}} \Theta(M,\hat{\theta}_{ij}^{lk}),
\end{align}
\normalsize
where we have defined $\hat{\theta}_{ij}^{lk}=\pi[\textrm{sin}(\hat{\theta}_{ij}) - \textrm{sin}(\hat{\theta}_{lk})]$ and $\Theta(M,\hat{\theta}_{ij}^{lk})=\nicefrac{\sin(M\hat{\theta}_{ij}^{lk}/2)}{\sin(\hat{\theta}_{ij}^{lk}/2)} e^{j\hat{\theta}_{ij}^{lk}(\frac{M-1}{2})}$. We highlight that $\mathbf{\hat{\bar{g}}}_{lkl}^H\mathbf{\hat{\bar{g}}}_{lkl}$ is a special case of \eqref{app3} when $i=l$ and $j=k$. Utilizing this observation we obtain $\mathbf{\hat{\bar{g}}}_{lkl}^H\mathbf{\hat{\bar{g}}}_{lkl} = \nicefrac{M\hat{\alpha}_{lkl}K_{lkl}}{\left(1+K_{lkl}\right)}$. Normalizing \eqref{app3} by $\mathbf{\hat{\bar{g}}}_{lkl}^H\mathbf{\hat{\bar{g}}}_{lkl}$, we obtain the expression for the LOS interference as
\small
\begin{align}\label{app5}
   \delta_{ij}^{lk} &= \frac{\hat{\alpha}_{ijl}K_{ijl}\left(1+K_{lkl}\right)} {\hat{\alpha}_{lkl}K_{lkl}\left(1+K_{ijl}\right)}\frac{|\Theta(M,\hat{\theta}_{ij}^{lk})|^2}{M^2},
\end{align}
\normalsize
where $\delta_{ij}^{lk} = |\nicefrac{\mathbf{\hat{\bar{g}}}_{lkl}^H\mathbf{\hat{\bar{g}}}_{ijl}}{\mathbf{\hat{\bar{g}}}_{lkl}^H\mathbf{\hat{\bar{g}}}_{lkl} }|^2$ represents the LOS interference between $\textrm{U}_{ij}$ and $\textrm{U}_{lk}$.

\end{document}